\documentclass[aps,preprint, superscriptaddress,nofootinbib]{revtex4}
\usepackage{amsmath}
\usepackage{graphicx}
\usepackage{subfigure}
\usepackage{amssymb}
\usepackage{xcolor}
\usepackage{multirow}
\usepackage{cancel}
\usepackage{color}
\usepackage{epsfig}
\usepackage{ulem}
\usepackage{listings}
\usepackage{xcolor}
\usepackage{float}
\usepackage{slashed}

\usepackage{tikz}
\usepackage{pgffor}							

\usepackage[colorlinks,linkcolor=black,citecolor=blue]{hyperref}
\usepackage{amsmath}
\usepackage{wrapfig}

\newcommand{\be}{\begin{equation}}
\newcommand{\ee}{\end{equation}}
\newcommand{\beq}{\begin{equation}}
\newcommand{\eeq}{\end{equation}}
\newcommand{\bea}{\begin{eqnarray}}
\newcommand{\eea}{\end{eqnarray}}
\newcommand{\besp}{\begin{equation}\begin{split}}
\newcommand{\eesp}{\end{split}\end{equation}}

\newcommand{\Dfbd}{\mathord{\buildrel{\lower3pt\hbox{$\scriptscriptstyle\leftrightarrow$}}\over {D}_{\mu}}}

\def\0{\textbf{0}}
\def\1{\textbf{1}}
\def\2{\textbf{2}}
\def\3{\textbf{3}}
\def\4{\textbf{4}}
\def\5{\textbf{5}}
\def\6{\textbf{6}}
\def\7{\textbf{7}}
\def\8{\textbf{8}}
\def\9{\textbf{9}}

\begin{document}
\title{Direct detection of Higgs portal for light self-interacting dark matter}

\author{Wu-Long Xu}
\email{wlxu@cdtu.edu.cn}
\affiliation{School of Electronic Engineering, Chengdu Technological University, Chengdu 611730, P. R. China}
\author{Jin Min Yang}
\email{jmyang@itp.ac.cn}
\affiliation{Institute of Theoretical Physics, Chinese Academy of Sciences, Beijing 100190, P. R. China}
\affiliation{School  of Physics, Henan Normal University, Xinxiang 453007, P. R. China}

\author{Bin Zhu}
\email{zhubin@mail.nankai.edu.cn}
\affiliation{Department of Physics, Yantai University, Yantai 264005,  P. R. China}

\begin{abstract}	
Self-interacting dark matter (SIDM) can address the small-scale anomalies and previous researches focused on such a SIDM heavier than GeV, for which the self-scattering cross-section is in the quantum resonance region and has a non-trivial velocity dependence. For a SIDM lighter than GeV, the self-scattering cross-section falls within the Born region. In this work, considering the constraints from CMB, BBN and the DM relic density, we investigate the direct detection of the Higgs portal for a sub-GeV SIDM with a scalar mediator. For this end, we consider two approaches : one is the cosmic-ray accelerated dark matter (CRDM) scattering off the nucleon, the other is the electron recoil caused by the halo dark matter. We present direct detection limits for the parameter space of light SIDM and scalar mediator.  
We find that the detectability in either approach needs a sizable mediator-Higgs mixing angle ($\sin\theta$) which is larger than one for the CRDM approach and larger than $10^{-3}$ for the electron recoil approach. While the former case cannot be realized in the Higgs-portal light SIDM model with a scalar mediator, the latter case may also be constrained by some astrophysical observations or beam dump experiment. Anyway, even if other constraints are quite stringent, the direct detection may provide independent limits for such a sub-GeV SIDM.  
  
\end{abstract}

\maketitle

\tableofcontents

\section{Introduction}\label{sec1}
Dark matter (DM) is the most mysterious substance and constitutes a major part of the 
universe ~\cite{Planck:2018vyg}. 
Currently, DM is known to only have gravitational interactions, and most evidences for its existence come from astronomical and cosmological observations~\cite{Ostriker:1973uit, Randall:2008ppe}. From the perspective of particle physics, if DM is a fundamental particle, it cannot be a component of the spectrum of the Standard Model (SM). Many new physics models that include DM particles have been proposed, such as the majoron DM model~\cite{Queiroz:2014yna}, the Bose–Einstein condensation DM models~\cite{Maurya:2024zao}, the self-interacting DM models~\cite{Arido:2024hfk},  and the supersymmetric singlet model~\cite{Wang:2022lxn}. The candidates for DM can be various, ranging widely from ultra-light DM with an energy scale of $10^{-22} \rm eV$ to primordial black holes with extremely high solar masses~\cite{Lin:2019uvt}, such as axion~\cite{Duffy:2009ig,Sikivie:2009fv}, nonthermal supermassive DM~\cite{Chung:1998ua,Kuzmin:1998kk}, non-topological solitons~\cite{Macpherson:1994wf,Frieman:1988ut} and primordial black holes~\cite{Kesden:2011ij}. One of the most popular candidates for DM is the weakly interacting massive particle (WIMP)~\cite{Buchmueller:2017qhf,Jungman:1995df,Lee:1977ua}, which can naturally give the DM relic density and detectable signatures at colliders. However, no clear signals of DM have been observed in detection experiments so far~\cite{GAMBIT:2017zdo,XENON:2024wpa,LUX:2022vee,PandaX:2023xgl,Elor:2021swj}. 

Moreover, DM plays a significant role in the evolution of the universe. The predictions of the $\Lambda\rm CDM$ model with cold collisionless DM are consistent with observations of large-scale structure of the universe. However, it faces several issues at small scales, such as the core-cusp problem~\cite{Salucci:2007tm,Burkert:1995yz}, missing satellites problem~\cite{Moore:1999nt,Kauffmann:1993gv}, the too big to fail problem~\cite{Boylan-Kolchin:2011qkt}, and the diversity problem~\cite{Oman:2015xda}. These small-scale anomalies can be addressed by hypothesizing self-interacting DM (SIDM) scenarios, with a self-scattering cross-section per unit mass within the range $0.1~\rm cm^2/g<\sigma/m_{\chi}<10~ cm^2/g$ ~\cite{Wang:2022lxn,Wang:2022akn,Spergel:1999mh,Chu:2024rrv,Chu:2018fzy,Colquhoun:2020adl,Tsai:2020vpi,Tulin:2012wi,Chu:2019awd,Wang:2014kja,Zhu:2021pad,Zhang:2016dck}. 

The simplest scenario of SIDM is a DM particle interacting with a light scalar or vector mediator particle~\cite{Tulin:2017ara}. However, the vector mediator scenario has been ruled out by the cosmic microwave background (CMB) and indirect detections of DM~\cite{Bringmann:2016din}. To relax these constraints, the possible solutions include considering the freeze-in mechanism of DM~\cite{Bernal:2015ova}, allowing the vector mediator particles to annihilate into lighter degrees of freedom~\cite{Duerr:2018mbd}, employing the p-wave annihilation of DM~\cite{Chu:2016pew,Kahlhoefer:2017umn}, considering the dark sector and visible sector not to reach a thermal equilibrium, having a temperature ratio not equal to one~\cite{Bernal:2019uqr,Zhu:2021pad}, and altering the type of portal interactions, such as the Higgs portal~\cite{Krnjaic:2015mbs} or the neutrino portal~\cite{Zhang:2023mcv,Duch:2019vjg,Kelly:2021mcd}. Additionally, the candidates for SIDM could also be dark atoms~\cite{Cline:2021itd}, nucleons~\cite{Gresham:2017zqi,Gelmini:2002ez}, glueballs~\cite{Carenza:2023eua,Carenza:2022pjd}, and bound states~\cite{Laha:2013gva,An:2015pva}. Also, some recent studies have investigated the realization of SIDM by finite-size DM~\cite{Chu:2018faw}, referred to as puffy DM, which suggests that the velocity dependence of $\sigma/m_{\chi}$ arises from the size of DM particle. A further study~\cite{Wang:2021tjf} finds that the dark strong interactions associated with puffy DM particles may dominate the scattering cross-section. Even neglecting the dark strong interactions, the study in~\cite{Wang:2023xii} finds that to obtain a more accurate self-scattering cross-section for puffy DM, a partial wave method should be used to solve the Schrödinger equation. This indicates that the partial wave method reveals a dependence on the ratio of $R_{\chi}$ (the radius of DM) to $m_{\phi}$ (the range), identifying three regions for the self-scattering cross-section of puffy DM: the effective Born region, the resonance region, and the classical region~\cite{Wang:2023wbw}.
 
 In light of such rich phenomena associated with SIDM, in this work we intend to study the direct detection of light SIDM through the Higgs portal. We consider the scenario with p-wave DM annihilation via a scalar mediator since it can relax the CMB constraints.  This scalar mediator can mix with the Higgs boson and then we can use the Higgs portal to couple the SM with the dark sector. This SIDM scenario with the Higgs portal can be probed by direct detection experiments and there have been a lot of relevant studies, most of which, however, focus on energies above the GeV scale~\cite{Kaplinghat:2013yxa,PandaX-II:2021lap,Jia:2019qcs,Duerr:2018mbd,Kahlhoefer:2017umn,Bringmann:2016din}. Note that the absence of clear signals in DM detections has led to an exploration for light sub-GeV DM, prompting upgrades and developments of numerous detection techniques~\cite{Campbell-Deem:2022fqm,Knapen:2020aky,Hochberg:2015pha,Ge:2020yuf,Alvey:2019zaa,Ema:2018bih}. Therefore, we will  investigate the direct detections of a light SIDM, considering the cosmic ray-accelerated DM (CRDM) ~\cite{Bringmann:2018cvk,Bondarenko:2019vrb,Wang:2021nbf,Xia:2021vbz,Wang:2023wrx} scattering off the nucleon and the electron recoil from the halo DM~\cite{Wu:2022jln,Baxter:2019pnz,Essig:2017kqs,XENON:2019gfn,Flambaum:2020xxo,Wang:2023xgm}. Further, although the non-trivial velocity dependence of the DM self-scattering cross-section in the resonance region has attracted much attention, the Born region for a light SIDM can also be used to address the small-scale problems. Thus, we will first consider fixing the dark sector coupling constant at $g_{\chi}\approx0.01m_{\chi}/270~\rm GeV$ (this approximation is adopted in~\cite{Alvarez:2019nwt,Boehm:2003hm,Pospelov:2007mp,Feng:2008ya,Feng:2008mu}) to satisfy the constraints of DM relic density. For the decay and lifetime constraints on the scalar mediator from CMB and Big Bang Nucleosynthesis (BBN) observations~\cite{Masi:2015fca,Slatyer:2015jla,Hufnagel:2018bjp,Depta:2019lbe,Poulin:2015opa,Arbey:2011nf}, we will use the results in ~\cite{Ding:2021zzg,Berger:2016vxi,Krnjaic:2015mbs}. By considering the detections of a light SIDM via CRDM and the electron recoil,  we aim to probe the un-explored regions for the mediator and its mixing with the Higgs boson. 

This paper is organized as follows. In Sec. II, we present a Higgs-portal SIDM model with scalar mediator and give the DM-nucleon and DM-electron interactions. In Sec. III, the CMB and BBN constraints on this model are described.  In Sec. IV, we constrain the light SIDM by considering the CRDM and the electron recoil from the halo DM.  Sec. V gives our conclusions.

\section{The Higgs portal SIDM model }\label{sec2}
This model is a simple extension of the SM, where an additional real scalar mediator $\Phi$ and a Dirac fermion $\chi$ are introduced. These two particles constitute the so-called dark sector, which communicates with the SM via the Higgs portal. A discrete $\rm Z_2$ symmetry is imposed to make the fermion $\chi$ stable to serve as the DM. The total Lagrangian including the renormalizable interaction between the dark sector and the visible SM sector is given by \cite{Krnjaic:2015mbs,Matsumoto:2018acr,Krnjaic:2015mbs}
\be\label{eq1}
\mathcal{L}=\mathcal{L_{\rm SM}}+\frac{1}{2}\bar{\chi}(i\partial\!\!\!/-m_{\chi})\chi+\frac{1}{2}\partial_{\mu}\Phi\partial^{\mu}\Phi-g_{\chi}\Phi\bar{\chi}\chi-V(\Phi,H),
\ee
where $\mathcal{L_{\rm SM}}$ is the SM Lagrangian involving the Higgs potential 
\be
V_H=\frac{\mu_1}{2}|H|^2+\frac{\mu}{4}|H|^2 ,
\ee
with $H$ being  the SM Higgs doublet.  The scalar potential $V(\Phi,H)$ is given by  
\be\label{eq2}
V(\Phi,H)=\frac{\lambda_1}{2}\Phi^2+\frac{\lambda_2}{4}\Phi^4+\frac{\lambda_{\Phi H}}{2}\Phi^2|H|^2.
\ee
When the electroweak symmetry is breaking and the unitary gauge is taken, the scalar fields are parameterized as 
\be\label{eq3}
H=\frac{1}{\sqrt{2}}
\begin{pmatrix}
	0\\
	v+h'
\end{pmatrix}\quad\quad\quad
\Phi=\frac{1}{\sqrt{2}}\left(u+\phi'\right),
\ee
where $v=246~\rm GeV$ is the vacuum expectation values of $H$. With the fields in eq.(\ref{eq3}),  we can get the mass terms
$m^2_{h'}$ and  $m^2_{\phi'}$ and the mixing term $m_{h'\phi'}^2$. Then we can obtain the physical fields $h$ and $\phi$: 
\be\label{eq4}
\mathcal{L}\supset -\frac{1}{\sqrt{2}}\left(h',\phi'\right)
\begin{pmatrix}
m^2_{h'}\quad\quad	m_{h'\phi'}^2\\
	m_{h'\phi'}^2 \quad\quad  m^2_{\phi'}
\end{pmatrix}
\begin{pmatrix}
h'\\
\phi'
\end{pmatrix}
=-\frac{1}{\sqrt{2}}\left(h,\phi\right)
\begin{pmatrix}
	m^2_h\quad\quad	0\\
	0 \quad\quad  m^2_{\phi}
\end{pmatrix}
\begin{pmatrix}
	h\\
	\phi
\end{pmatrix}.
\ee
with the rotation 
\be\label{eq5}
\begin{pmatrix}
	h\\
	\phi
\end{pmatrix}
=
\begin{pmatrix}
	\cos \theta\quad\quad	-\sin\theta\\
	\sin\theta \quad\quad  \cos\theta 
\end{pmatrix}
\begin{pmatrix}
	h'\\
	\phi'
\end{pmatrix},
\ee
and the relations 
\be\label{eq6}
m^2_{h(\phi)}=\frac{1}{2}\left(m^2_{h'}+m^2_{\phi'}\pm \sqrt{(m^2_{h'}-m^2_{\phi'})^2+4m^4_{h'\phi'}}\right),  \quad\quad
\tan2\theta=-\frac{2m^2_{h'\phi'}}{m^2_{h'}-m^2_{\phi'}}.
\ee
In the following, the SM Higgs mass $m_h$ is fixed to $125~ \rm GeV$.  The coupling between the scalar $\phi$ and the SM fermion $f$ is expressed as 
\be\label{eq7}
y_{\bar{f}f\phi}=\frac{m_f\sin\theta}{v}.
\ee
In our numerical calculation, the input parameters are chosen to be 
\be\label{eq8}
\sin\theta, \quad m_{\chi},\quad m_{\phi}, \quad g_{\chi} 
\ee
Note that from a phenomenological perspective,  the mixing angle $\theta$ is usually required to be very small. However, for a very light SIDM to be detectable in direct detections, this mixing angle cannot be small, as shown in our following study. And our aim is to explore the allowed parameter space for $\sin\theta$ with a given dark matter mass $m_{\chi}$. 

For the direct detection of the Higgs-portal light DM scattering off the nucleon, the relevant DM-nucleon interaction can be written as~\cite{Alvey:2022pad}
\be\label{eq19}
\mathcal{L}_{\rm \chi-N}^{\rm scalar}\supset g_{\chi}\phi\bar{\chi}\chi+g_N\phi\bar{N}N,
\ee
where $N$ is a nucleon and its mass is denoted by the proton mass $m_p$. From Eq. (\ref{eq7}), it can be seen that   the coupling $g_N$ can be replaced by the coupling between mediator $\phi$ and the SM fermion:  
\be\label{eq20}
g_{\rm N}\approx\frac{m_p\sin\theta}{v}f_p\approx 1.1\times10^{-3}\times\sin\theta,
\ee
where $f_p\approx0.3$ is the effective nucleon coupling~\cite{Cline:2013gha}.

For the direct detection of the Higgs-portal light DM via electron recoil,  the relevant DM-electron interaction can be written as
\be \label{eq28}
\mathcal{L}_{\rm \chi-e}^{\rm scalar}\supset g_{\chi}\phi \bar{\chi}\chi -g_e\phi\bar{e}e,
\ee 
where the coupling $g_e$ is given by 
\be \label{eq29}
g_e\approx \frac{m_e\sin\theta}{v}=2.077\times 10^{-6}\times \sin\theta.
\ee 

\section{Cosmological and astrophysical constraints}\label{sec3}

\subsection{Dark matter self-interaction solving small-scale problems }\label{sec3.1}
In our Higgs-portal SIDM model, the scattering of two non-relativistic DM particles can be described by a Yukawa  potential
\be\label{eq9}
V(r)=\frac{\alpha_{\chi}}{r}e^{-m_{\phi}r},
\ee
where $\alpha_{\chi}=g_{\chi}^2/4\pi$ and, due to the scalar mediator, their interaction is attractive purely. The scattering cross section is dependent on the DM relative velocity $v$. Here the transfer cross section $\sigma_T$  weighted by $(1-\cos\theta )$ is adopted ~\cite{Tulin:2013teo}
\be\label{eq10}
\sigma_T=\int d\Omega (1-\cos\theta)\frac{d\sigma}{d\Omega}.
\ee
Generally, the self-scattering cross section can be divided into  Born region ($\alpha_{\chi}m_{\chi}/m_{\phi}\ll 1$) where $\sigma_T$ can be calculated by the perturbative approach, the classical region ($m_{\chi}v/m_{\phi}\gg1$) which belongs to non-perturbative regime and its cross section $\sigma_T$  also can be obtained via the analytic formulae,   and the  residual resonant region. In order to solve the small-scale anomalies,  we require $\sigma_T/m_{\chi}\simeq 0.1-10~\rm cm^2g^{-1}$.  Such a large velocity-dependent cross section usually needs a long-range force, which means the existence of a light mediator (another source of velocity dependence for $\sigma_T/m_{\chi}$ is considered in ~\cite{Chu:2018faw}). Meanwhile, the strongly coupling resonant region of self-scattering is also interesting, due to its non-trivial velocity dependence. 

The coupling $\alpha_{\chi}$ is fixed by the freeze-out mechanism of DM production in the early universe, which is  $0.01~m_{\chi}/270 ~\rm GeV$. For the direct detection of a light SIDM,  the Born regime for $\sigma_T/m_{\chi}$ ($m_{\chi}\leq 10 ~\rm GeV, m_{\phi}\leq 10~\rm MeV$) should be explored.

Here we just give the formula of momentum-transfer cross section in the Born limit (for the details of other regimes for $\sigma_T$, see ~\cite{Tulin:2013teo})
\be\label{eq11}
\sigma_T^{\rm Born}=\frac{8\pi\alpha_{\chi}^2}{m^2_{\chi}v^4}\left(\log(1
+m_{\chi}v^2/m^2_{\phi})-\frac{m^2_{\chi}v^2}{m^2_{\phi}+m^2_{\phi}v^2}\right).
\ee
In the astrophysical observations of small scale structures, the DM relative velocity follows the Maxwell-Boltzmann distribution. Thus, we use the velocity-averaged transfer cross section 
\be\label{eq12}
\left<\sigma_Tv\right>=\int \frac{32v^2e^{-4v^2/\pi\left<v\right>^2}}{\pi^2\left<v\right>^3}\sigma_Tvdv,
\ee
where $\left<v\right>=v_0$ is the characteristic velocity of different size halos. As shown in the right panel of Fig. \ref{fig2}, the blue region is for dwarf scale $(v_0/c=10$), red zone for MW scale $(v_0/c=200$) and green region for cluster scale $(v_0/c=10^3$).  The cross section $\left<\sigma_Tv\right>/m_{\chi}$ is constrained in a range $(0.1-10~\rm cm^2/g\times km/s)$.  In the overlap region of the parameter space $(m_{\phi},m_{\chi})$ the SIDM can solve the small-scale structure anomalies. Therefore, we wonder whether or not the direct detection has sensitivity to this overlap region. 

\subsection{CMB and BBN constraints}\label{sec3.2}
During the recombination stage of the early universe, DM annihilation may modify the temperature and polarization power spectra of CMB~\cite{Chen:2003gz,Padmanabhan:2005es}.  The DM annihilation cross section can be bounded from the measurement of CMB at redshift $z\sim 1100$ and the velocity of DM is smaller than $10^{-7}$. Thus, according to the recent Planck data, the  annihilation cross section of DM is constrained. At $95\%$  C.L., the result is expressed as ~\cite{Ding:2021zzg,Planck:2018vyg}
\be\label{eq13}
f_{\mathrm{eff}}\langle\sigma_{\rm ann}v_{\rm rec}\rangle/m_{\chi}\leq3.2\times10^{-28}\rm cm^{3}s^{-1}GeV^{-1},
\ee
where the efficiency factor $f_{\rm eff}$ denotes the fraction of the released energy ending up in photons or electrons ($f_{\rm eff}=0.16$ is taken for $4\mu$ annihilation channel~\cite{Madhavacheril:2013cna}). For the annihilation cross section, the Sommerfeld enhancement effect should be considered because of  the very low velocity of the DM particle. Note that during the freeze-out of DM, its velocity is relativistic and is about $0.3c$. The corresponding annihilation rate is $\langle\sigma v\rangle^{\rm ann}_{\rm therm}\sim 3\times 10^{-26} ~\rm cm^3s^{-1}$ where $\sigma v \propto 1/v$.  At the recombination stage, the annihilation cross section can be larger than the freeze-out value by one order.  In the following, we study the Sommerfeld enhancement constrained with CMB.

First, we start with the DM  relic density. The annihilation process is $\bar{\chi}\chi\rightarrow \phi\phi$.  For comparison with the vector mediator case, we just consider the tree-level situation where the mediator $\phi$ can be either a dark vector field or a dark scalar field. Thus,  the annihilation cross sections  are
\begin{align}\label{eq14}
(\sigma_{\rm ann}v) & = \begin{cases}
\frac{\pi\alpha_{\chi}^{2}}{m_{\chi}^{2}}\sqrt{1-\frac{m_{\phi}^{2}}{m_{\chi}^{2}}} & \quad {\rm vector ~mediator\quad(s~ wave ~case)}\\
\hspace{5cm}\  & \ \\[-6.mm]
\frac{3\pi\alpha_{\chi}^{2}}{4m_{\chi}^{2}}v^{2}\sqrt{1-\frac{m_{\phi}^{2}}{m_{\chi}^{2}}}&\quad  {\rm scalar~ mediator\quad(p ~wave~ case)}.
\end{cases}
\end{align}
Thus, the annihilation cross section with Sommerfeld enhancement is $\sigma_{\rm ann}v=S\times(\sigma_{\rm ann}v)^{\rm tree}$ with $S$ being the Sommerfeld enhancement factor. The thermally averaged annihilation cross section, in which the velocity distribution function $f(v)$ of DM is the Maxwell-Boltzmann distribution~\cite{Feng:2010zp}, is given by 
\be\label{eq15}
\langle\sigma_{\rm ann}v\rangle=\frac{x^{3/2}}{2\sqrt{\pi}}\int dv\frac{v^{2}}{2}e^{-\frac{xv^{2}}{4}}S(\sigma_{\rm ann}v)^{\rm tree},
\ee
where  $x$ can be replaced by the most probable velocity $v_0$, namely, $v_0=\sqrt{2T_{\chi}/m_{\chi}}=\sqrt{2/x_{\chi}}$. The Sommerfeld enhancement factors for s-wave and p-wave annihilations are 
\be\label{eq16}
S_{s}=\frac{\pi}{a}\frac{\sinh(2\pi ac)}{\cosh(2\pi ac)-\cos(2\pi\sqrt{c-(ac)^{2}})} \quad\quad S_{p}=\frac{(c-1)^{2}+4(ac)^{2}}{1+4(ac)^{2}}S_{s},
\ee
where the parameters $a$ and $c$ are defined as $\frac{v}{2\alpha_{\chi}}$and $ \frac{6\alpha_{\chi}m_{\chi}}{\pi^{2}m_{\phi}}$, respectively~\cite{Tulin:2013teo}. 

\begin{figure}[ht]
	\centering
	\includegraphics[width=6.5cm]{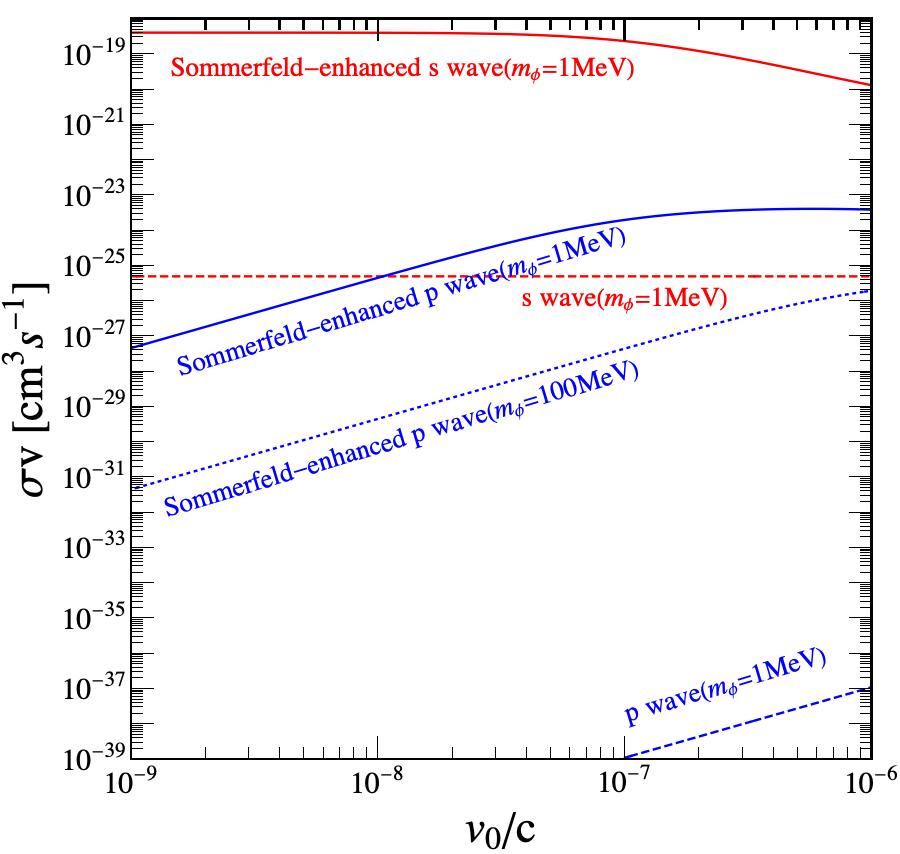}
	\includegraphics[width=6.3cm]{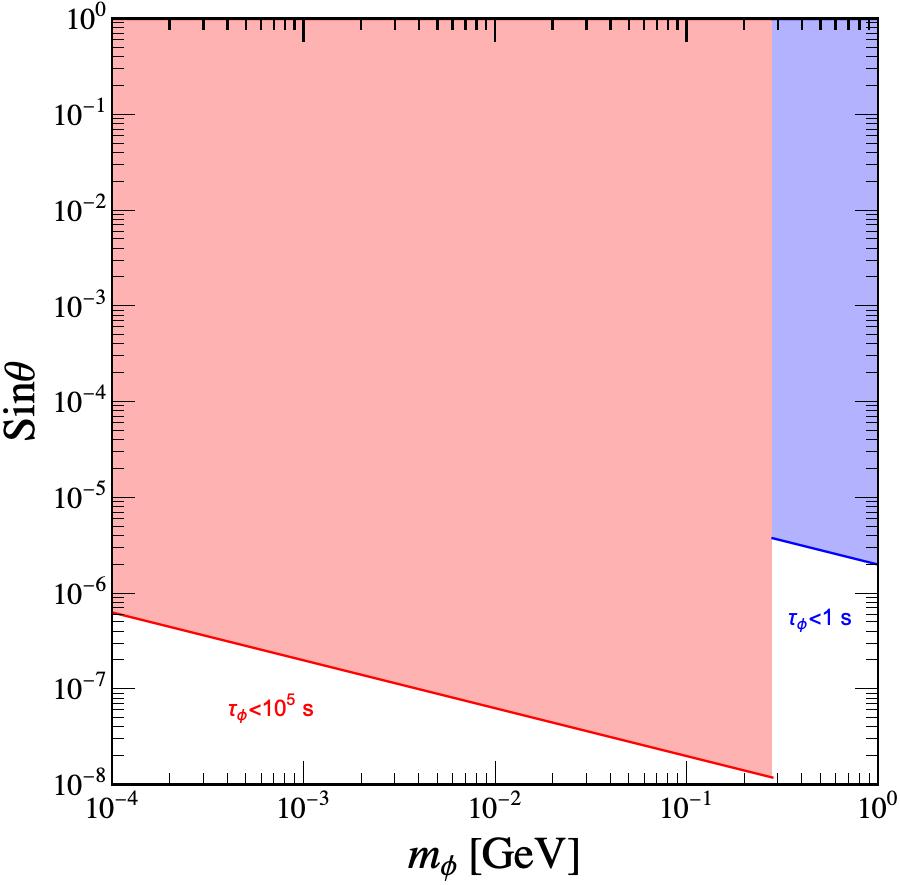}
 \vspace{-.3cm}
	\caption{Left panel shows the annihilation cross sections for s-wave and p-wave as a function of the DM velocity for the CMB epoch, where the mass of DM is $m_{\chi}=1~\rm TeV$.
		Right panel shows the constraint on the mediator $\phi$ decay from BBN (the shaded regions are allowed).}
	\label{fig1}
\end{figure}

After the DM freeze out, its velocity will decrease. Specially, during the recombination epoch, the relative velocity of the DM is about $10^{-8}-10^{-6}c$. According to Eq. (\ref{eq15}), the DM annihilation cross section in  different stages of cosmological evolution can be obtained. In the  left panel of  Fig. \ref{fig1}, a comparison of annihilation cross sections for s-wave and p-wave is given for the CMB epoch. It shows that the Sommerfeld enhanced s-wave annihilation cross section is very large (much larger than the constant freeze-out cross section $<\sigma v>^{\rm ann}_{\rm therm}$) so that the s-wave annihilation in the vector mediator scenario is stringently constrained by the CMB, as shown in Eq. (\ref{eq13}).  In contrast, the Sommerfeld enhanced p-wave annihilation cross section in the scalar mediator scenario is  rather small. So the Higgs-portal light SIDM with a scalar mediator considered in this work can survive the CMB constraint.

In addition, the BBN can also give some constraints on the Higgs-portal DM model. Because the dark particles can modify the light nuclei abundances when the Hubble constant and entropy injection are re-corrected by the mediator decay. And these constraints are related with the abundance and lifetime of this scalar mediator.  To avoid overclosing the universe, the constraint from the requirement of the mediator $\phi$ decay before BBN can be expressed as ~\cite{Berger:2016vxi,Krnjaic:2015mbs}
\be\label{eq17}
\begin{split}
\tau_{\phi}&\leq 1~\rm s  \quad  \quad   \quad {\rm if ~m_{\phi}\geq 2m_{\pi} }\\
\tau_{\phi}&\leq10^5~ \rm s  \quad  \quad  {\rm~if~ m_{\phi}\leq 2m_{\pi} }.
\end{split}
\ee
For the light mediator decay, the dominant decay channel is $\phi\rightarrow e^+e^-$. The mediator decay rate and lifetime from Eq. (\ref{eq7}) are 
\be\label{eq18}
\begin{split}
\Gamma_{\phi}&=\frac{\sin^2\theta m^2_em_{\phi}}{8\pi v^2}\\
\tau_{\phi}&\approx \left(\frac{\sin\theta}{10^{-5}}\right)^{-2}\left(\frac{10~\rm MeV}{m_{\phi}}\right)\times 4 ~\rm s
\end{split}
\ee
The corresponding results are shown in the right panel of Fig. \ref{fig1}. We see that the mixing angle $\sin\theta$ must be larger than $1\times 10^{-8}$, which gives a lower limit $2.077\times 10^{-14}$ on the coupling $y_{ff\phi}$. And when the mediator is lighter, this mixing angle becomes larger, which should be also constrained by other astrophysical constraints from  SN1987a, HB and RG stars (for a detailed discussion, see ~\cite{Raffelt:1996wa}) and the beam dump constraints~\cite{Liu:2016qwd}.

\section{Direct detection}\label{sec4}
In addition to the constraints from astrophysical observations,  the direct detection can give an independent constraint on the Higgs-portal SIDM. So far the focus of direct detection has been mainly on a DM with mass above $\rm GeV$. On the one hand, for a DM heavier than $\rm GeV$, its self-scattering has rich dynamic effect. The scattering cross section with a light mediator is in the quantum resonant region and has an amazing velocity dependence. On the other hand, traditional DM direct detection technology cannot explore a DM below the $\rm GeV$ scale. In the following,  we give two complementary ways for the  direct detection of the Higgs-portal light SIDM with a scalar mediator.

\subsection{Detection of cosmic ray-boosted dark matter}\label{sec4.1}
When the mass of DM is below $1~\rm GeV$, the  direct detection sensitivity for the halo DM is very low due to the tiny nuclear recoil energy. In this case, the cold halo DM particles may collider with high-energy galactic cosmic ray (CR) 
and thus get boosted.  Such CR-boosted light DM (CRDM) flux can be (semi-)relativistic. 

The differential scattering cross section describing the collision between a CR particle and a DM particle can be written as 
\bea\label{eq21}
\left(\frac{d\sigma_{\chi N}}{dT_{\chi}}\right)_{\rm{ scalar}}&=&\frac{A^2g_{N}^{2}g_{\chi}^{2}F^{2}(q^{2})(2m_{\chi}+T_{\chi})(2m_{N}^{2}+m_{\chi}T_{\chi})}{8\pi T_{i}(T_{i}+2m_{N})(m_{\phi}^{2}+2m_{\chi}T_{\chi})^{2}},
\eea
where $A$ is the atomic number and $F$ is  taken as the usual dipole nucleon form factor $F(q^2)=(1+q^2/\Lambda^2)^{-2}$ with $\Lambda=0.843 \rm ~GeV$ ~\cite{Alvey:2022pad}. $T_i$ and $T_{\chi}$ are the incoming and outgoing kinetic energies, respectively. 

\begin{figure}[ht]
	\centering
\includegraphics[width=14.5cm]{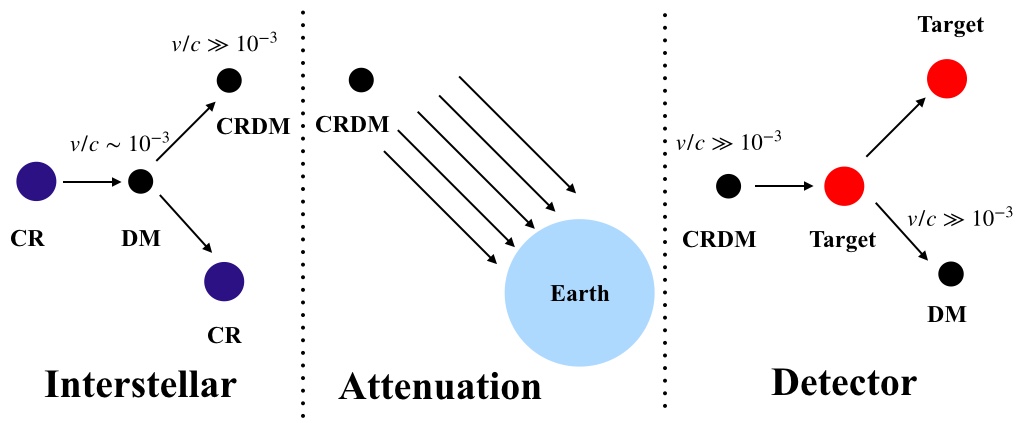}
	\caption{Left panel:DM boosted by CR; Middel penal: attenuation process;
		Right panel: CRDM scattering in detectors. }
	\label{fig21}
\end{figure}
The whole process of CRDM from being boosted to reaching detector includes the following three steps:

{\bf (1) DM boosted by CR:} As shown in the left picture of Fig.~\ref{fig21},  before scattering with the CR, the low-velocity background DM can be considered as at rest. Then DM particles get boosted by energetic CR. The differential relativistic  CRDM flux is given by  
\be\label{flux}
\frac{d\Phi_{\chi}}{dT_{\chi}}=D_{\rm eff}\frac{\rho_{\chi}^{\rm local}}{m_{\chi}}\sum_{i}\int_{T_{i}^{\rm min}}dT_{i}\left(\frac{d\sigma_{\chi i}}{dT_{\chi}}\right)_{\rm scalar}\frac{d\Phi_{i}^{\rm LIS}}{dT_{i}}~,
\ee
where the local DM halo density  $\rho_{\chi}=0.3~  \rm GeV cm^{-3}$,
the effective diffusion zone parameter $D_{\rm eff}=0.997~ \rm kpc$ as in ~\cite{Bringmann:2018cvk}.  The local interstellar CR spectrum (LIS) flux for $i={p,He}$ is denoted as $d\Phi_i^{\rm LIS}/dT_i$~\cite{Boschini:2017fxq,DellaTorre:2016jjf}. Given a CRDM energy $T_{\chi}$, the minimal incoming CR energy  $T_i^{\rm min}$ is obtained as 
\be\label{eq23}
T_i^{\rm min}=\left(\frac{T_{\chi}}{2}-m_i\right)\left[ 1\pm\sqrt{1+\frac{2T_{\chi}}{m_{\chi}}\frac{(m_i+m_{\chi})^2}{(2m_i-T_{\chi})^2}} ~\right].  
\ee

{\bf (2) Attenuation:}  To avoid background events, many detectors are located in the deep underground. As shown in the middle picture of Fig.~\ref{fig21}, the CRDM flux flows from the top of the atmosphere to the location of the detector and in this process the CRDM energy will be attenuated by dense matter of the earth. Thus, the degradation of CRDM energy as CRDM travels a distance $x$ can be estimated numerically~\cite{Starkman:1990nj,Emken:2018run}
\be \label{att1}
\frac{dT_{\chi}^{z}}{dz}=-\sum_{N}n_{N}\int_{0}^{T_{N}^{\rm max}}\frac{d\sigma_{\chi N}}{dT_{N}}T_{N}dT_{N}~,
\ee 
where $T_{\chi}^z$ denotes the CRDM energy located at the $z$-depth from the top of the atmosphere of the earth, $n_N$  is the average nuclei densities of the earth's elements provided by {\bf DarkSUSY 6.4.0} ~\cite{Bringmann:2018lay}, $T_N$  is the recoil energy of nucleus $N$ which includes the dominant elements  {O, Ca, C, Mg, Si, Al, K}. The differential cross section $d\sigma_{\chi N}/dT_N$ is calculated at the $z$-depth.  The attenuated CRDM flux located at the $z$-depth can be obtained via the primordial flux $d\Phi_{\chi}/dT_{\chi}$ 
\be \label{att2}
\frac{d\Phi_{\chi}}{dT_{\chi}^{z}}=\left(\frac{dT_{\chi}}{dT_{\chi}^{z}}\right)\frac{d\Phi_{\chi}}{dT_{\chi}} ~.
\ee

{\bf (3) CRDM scattering in detector:} As shown in the right picture of Fig.~\ref{fig21},  if  CRDM retains enough energy after attenuation, the remaining flux may produce detectable signals in the detector.   In the recoil energy range $T_1<T_N<T_2$, the recoil event rate per unit detector-particle mass is given by 
\be\label{rate}
R=\int_{T_{1}}^{T_{2}}\frac{1}{m_{N}}dT_{N}\int_{T_{\chi}^{\rm z,min}}^{\infty}dT_{\chi}^{z}\frac{d\sigma_{\chi N}}{dT_{N}}\frac{d\Phi_{\chi}}{dT_{\chi}^{z}}~.
\ee
The differential cross section can be obtained from Eq. (\ref{eq21})  via the substitutions $m_{\chi}\leftrightarrow m_N$, $T_i \rightarrow T_{\chi}$ and $T_{\chi} \rightarrow T_N$.  And it can be calculated as 
\bea\label{seccc}
	\left(\frac{d\sigma_{\chi N}}{dT_{N}}\right)_{{\rm {scalar}}}&=&\frac{A^{2}g_{N}^{2}g_{\chi s}^{2}F^{2}(q^{2})(2m_{N}+T_{N})(2m_{\chi}^{2}+m_{N}T_{N})}{8\pi T_{\chi}(T_{\chi}+2m_{N})(m_{\phi}^{2}+2m_{N}T_{N})^{2}}.
\eea
Note that the three differential cross sections for three scattering processes are expressed in different frames.  

According to the recent Xenon1T data, the nuclear recoil energy $T_{\rm Xe}\in [4.9,40.9]~\rm keV$~\cite{XENON:2018voc}. We modified the package {\bf DarkSUSY 6.4.0} where, for simplicity, we neglect the inelastic scattering during the attenuation process. By comparing Xenon1T experimental results with theoretical predictions, direct detection constraints on light CRDM can be obtained numerically.

\begin{figure}[ht]
	\centering
	\includegraphics[width=6.5cm]{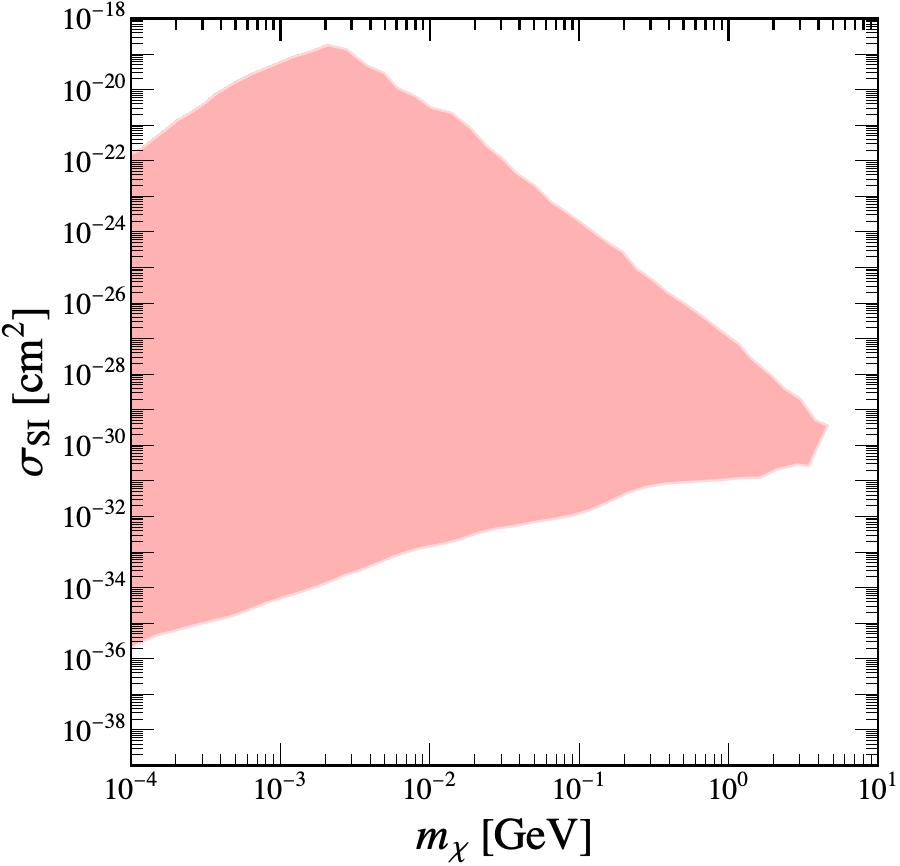}
	\includegraphics[width=6.3cm]{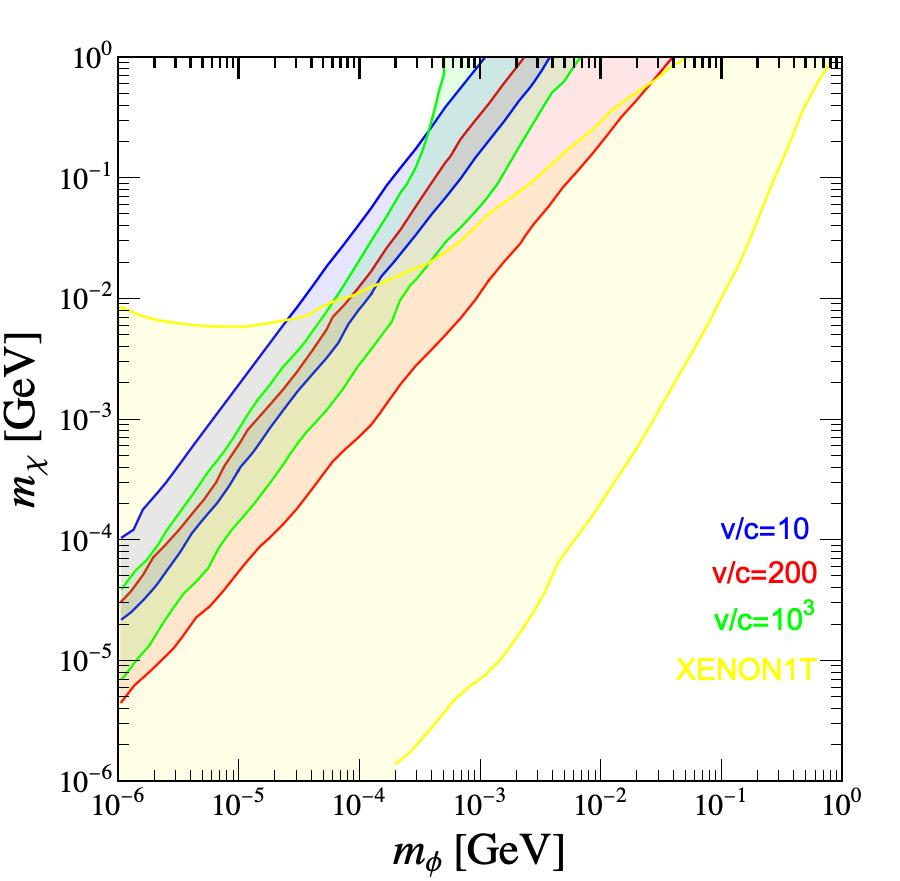}
 \vspace{-.3cm}
	\caption{Left panel: the XENON1T exclusion region (red) on the plane of the spin-independent CRDM-nucleon scattering cross section versus the CRDM mass.
		Right panel: each region sandwiched by a pair of curves (blue, red, green) is allowed by solving the small-scale problems, while the yellow region is excluded by XENON1T.}
	\label{fig2}
\end{figure}
First, the coupling $y_{\chi}$ is fixed by the DM relic density requirement. The coupling $y_N$ is replaced by the mixing angle $\sin\theta $ from Eq. (\ref{eq20})  and then constrained by the BBN from Eq. (\ref{eq17}). 
Following this, constraints on the spin-independent scattering cross section of CRDM  with nucleon can be derived from the Xenon1T data by examining the interaction between the relativistic DM flux and nucleon.

The left panel of Fig. \ref{fig2} shows 
the XENON1T exclusion region on the plane of the spin-independent CRDM-nucleon scattering cross section versus the CRDM mass, where the coupling $y_{\chi}$ is fixed by the DM relic density requirement and $y_N$ is constrained by the BBN  
We see that the limits on CRDM mass can reach up to $10~\rm GeV$ due to the form factor dependence of the cross section in the attenuation part, while the exclusion for the cross section is $10^{-19} ~\rm cm^2<\sigma_{\rm SI} < 10^{-36} ~\rm cm^2$ in the mass range $0.1 ~\rm MeV\geq m_{\chi}\leq 10~ GeV $.  
The right panel of Fig. \ref{fig2} shows the regions allowed by solving the small-scale problems, compared with the XENON1T exclusion region.
We see that the XENON1T exclusion can cover the most part of the parameter space allowed by solving the small-scale problems. 

\begin{figure}[ht]
	\centering
	\includegraphics[width=6.5cm]{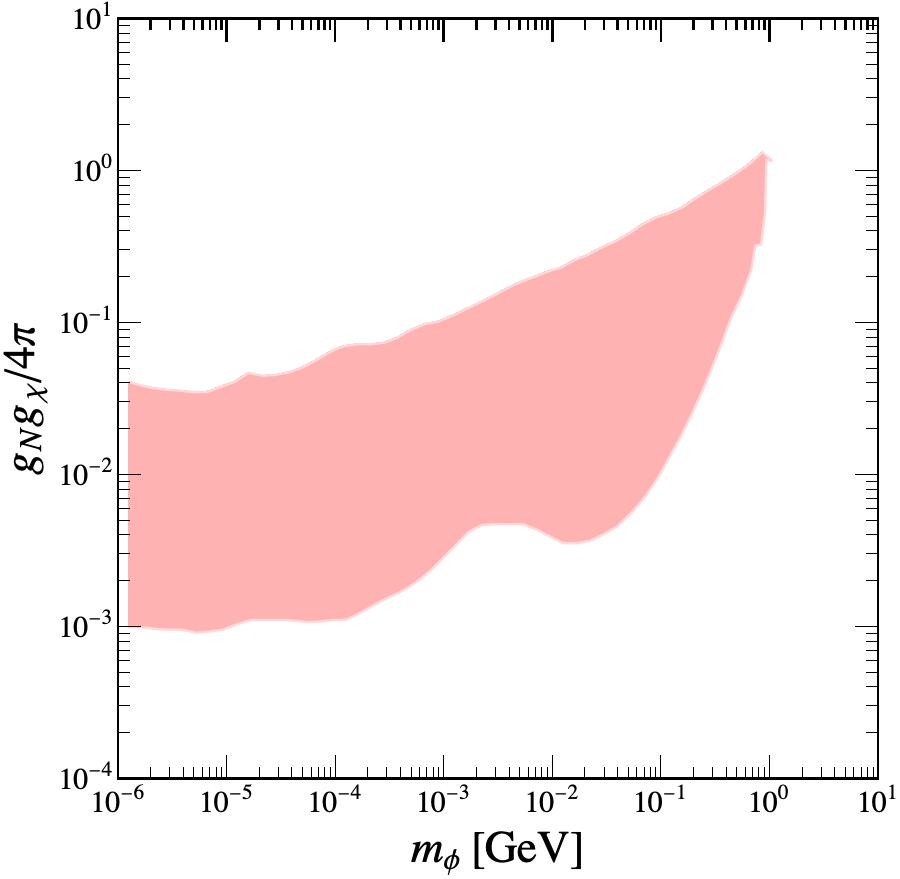}
	\includegraphics[width=6.5cm]{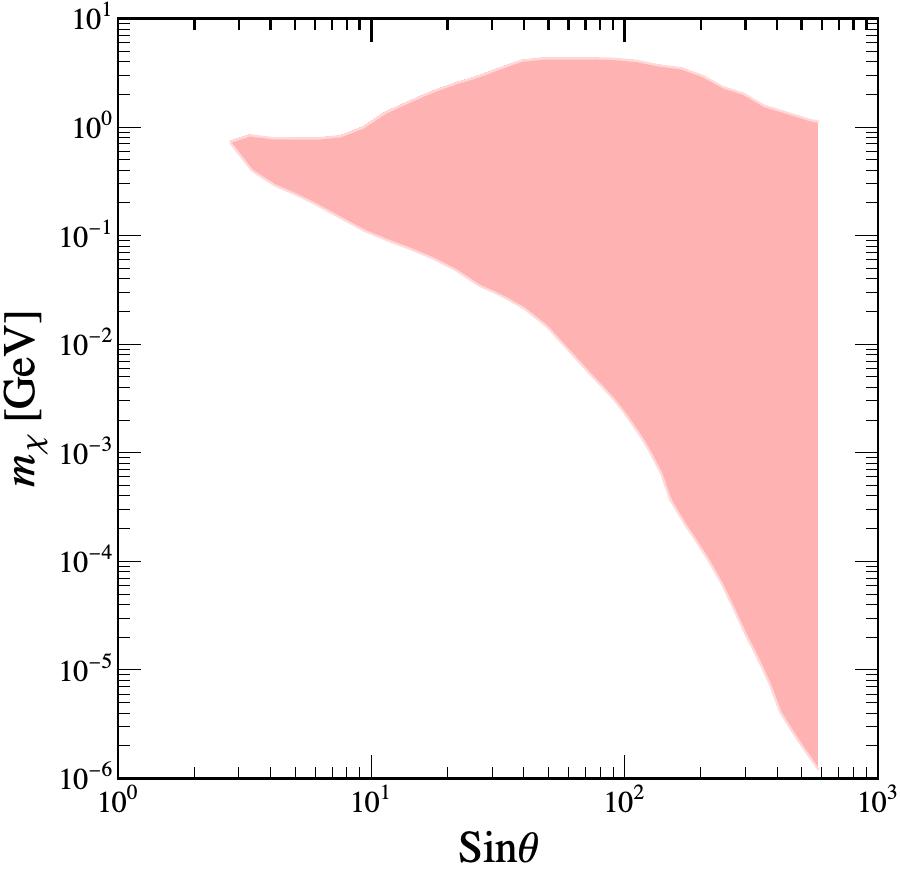}
  \vspace{-.5cm}
	\caption{The XENON1T exclusion regions (red) on the plane of $g_{\chi}g_{N}/4\pi$ versus the scalar mediator mass (left panel) and on the plane of CRDM mass versus $\sin\theta$ (right panel).}
	\label{fig3}
\end{figure}

Finally, we present the constraints on the sub-GeV scalar mediator coupling to the nucleon from the direct detection of CRDM in Xenon1T in the left panel of Fig. \ref{fig3}. It indicates that in the exclusion region the total coupling $g_{\chi}g_{N}/4\pi$ is within the perturbative regime, in comparison to the exclusion region in Fig. 3 of Ref. {~\cite{Bell:2023sdq}} where the scalar mediator gives a non-perturbative coupling. In our work, when the BBN and the relic density  constraints are considered, the non-perturbative coupling regime is excluded. However, we observe from the right panel of Fig. \ref{fig3} that in the exclusion region the mixing angle $\sin\theta$ is large than one which is not possible in our model. This means that the direct detection of CRDM needs a sizable mediator-nucleon coupling $g_N\sim m_p\sin\theta/v$ shown in Eq.(\ref{eq20}).

\subsection{Detection of halo dark matter via electron recoil}\label{sec4.2}
Now we turn to the direct detection of halo dark matter via electron recoil.
The DM-electron scattering cross section can be described by the reference cross section  $\bar{\sigma}_e$ and the momentum-dependent factor $F_{\rm DM}(q)$
\be \label{eq30}
\sigma_e=\bar{\sigma}_e\times |F_{\rm DM}(q^2)|^2,
\ee 
where the reference cross section is defined as 
\be \label{eq31}
\bar{\sigma}_e=\frac{g_{\chi}^2g^2_e}{4\pi}\frac{\mu^2_{\chi e}}{(m^2_{\phi}+\alpha^2m_e^2)^2},
\ee 
with $\mu_{\chi e}$ being the DM-electron reduced mass. The form factor can be written as 
\be \label{eq32}
F(q)=\frac{(m_{\phi}{}^{2}+q_{0}^{2})^{2}(4m_{e}^{2}+q^{2})(4m_{\chi}^{2}+q^{2})}{(m_{\phi}{}^{2}+q^{2})^{2}(4m_{e}^{2}+q_{0}^{2})(4m_{\chi}^{2}+q_{0}^{2})}.
\ee 

For calculating the DM-electron scattering rate in the liquid Xenon, the details can be found in Refs. ~\cite{Essig:2012yx,Essig:2017kqs,Wang:2023xgm}. The velocity-averaged differential cross-section of DM scattering off electron in the $(n, l)$ shell is given by 
\be\label{eq33}
\frac{d\left<\sigma_{\rm ion}^{nl}\right>}{d\ln E_{R}}=\frac{\bar{\sigma}_{e}}{8\mu_{\chi e}^{2}}\int qdq|f_{\rm ion}^{nl}(k',q)|^{2}|F_{\rm DM}(q)|^{2}\eta(v_{\rm min}),
\ee
where the momentum $k'=\sqrt{2m_eE_R}$, $E_R$ is the electron recoil energy  and $q$ denotes the momentum transfer.  The inverse mean speed $\eta(v_{\rm min})=\left<\frac{1}{v}\theta(v-v_{\rm min})\right>$ obeys the  Maxwell-Boltzmann velocity distribution of DM with the circular velocity $v_0=220~ \rm km/s$  and the escaped velocity $v_{\rm esc}=544~ \rm km/s$. The minimum velocity $v_{\rm min}$ of the DM particle required for the scattering is given by
\be\label{eq34}
v_{\rm min}=\frac{|E_b^{nl}|+E_R}{q}+\frac{q}{2m_{\chi}},
\ee 
with  $E_b$ being the binding energy.
The form factor $f_{\rm ion}^{nl}(k',q)$ for ionization of electron in the $(n,l)$ shell can be expressed as 
\be\label{eq35}
|f_{\rm ion}^{nl}(k',q)|^{2}=\frac{k'^{3}}{4\pi^{3}}\times2\underset{n,l,l'm'}{\sum}|\langle f|e^{i{\bf q}_{e}\cdot{\bf x}_{i}}|i\rangle|^{2}.
\ee 
The differential ionization rate is obtained as 
\be \label{eq36}
\frac{dR_{\rm ion}}{d\ln E_{R}}=N_{T}\frac{\rho_{\chi}}{m_{\chi}}\sum_{nl}\frac{d\left<\sigma_{\rm ion}^{nl}v\right>}{d\ln E_{R}},
\ee 
where $N_T$ is the number of target atoms and $\rho_{\chi}=0.4~\rm GeV/cm^3$ is the local matter density.

\begin{figure}[ht]
	\centering
	\includegraphics[width=6.5cm]{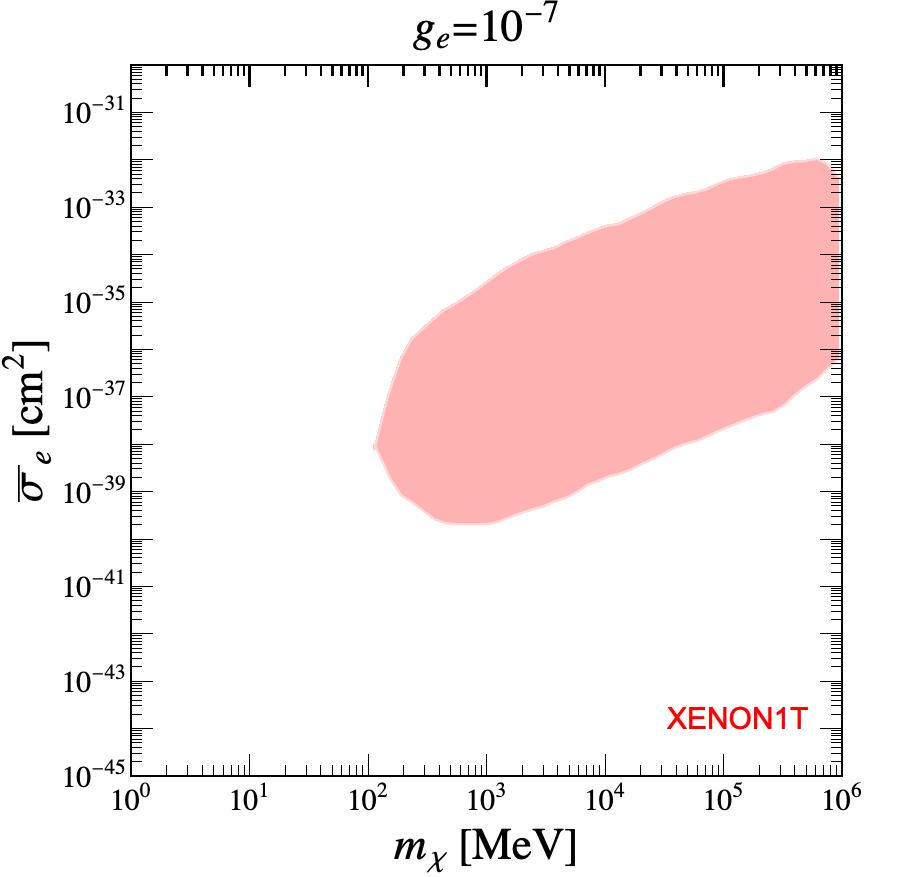}
	\includegraphics[width=6.3cm]{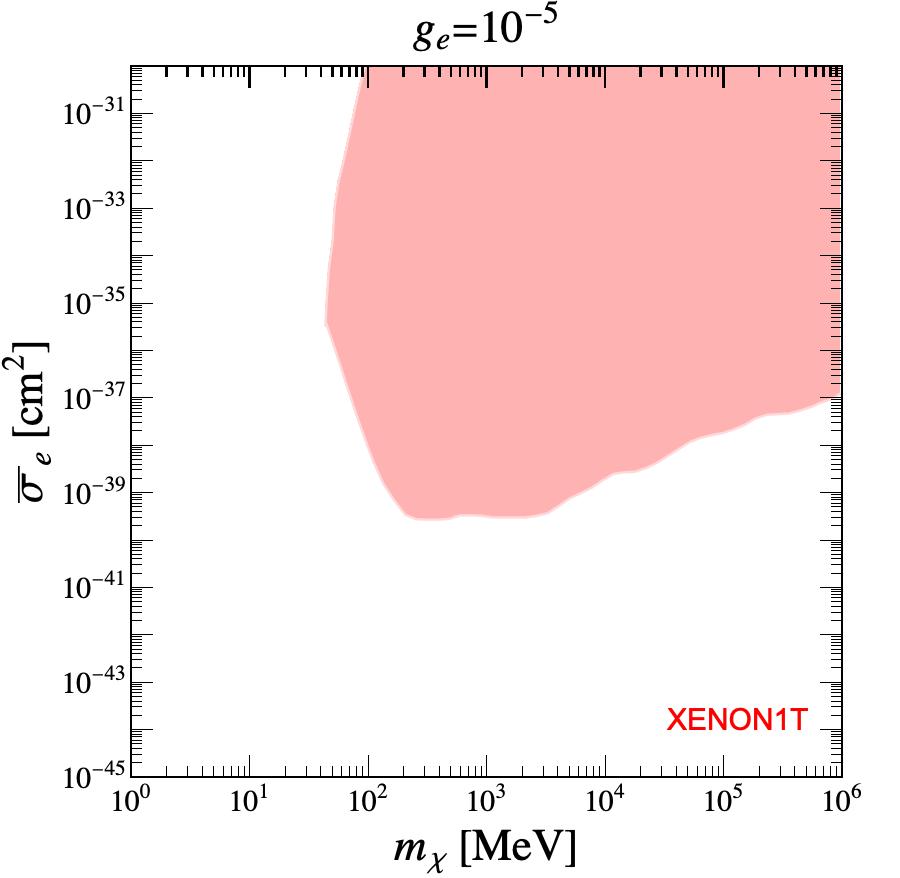}
  \vspace{-.5cm}
	\caption{ The XENON1T exclusion regions on the plane of DM-electron scattering cross-section versus the DM mass for the coupling $g_e=10^{-5}$ and $g_e=10^{-7}$.}
	\label{fig4}
\end{figure}

\begin{figure}[ht]
	\centering
	\includegraphics[width=6.5cm]{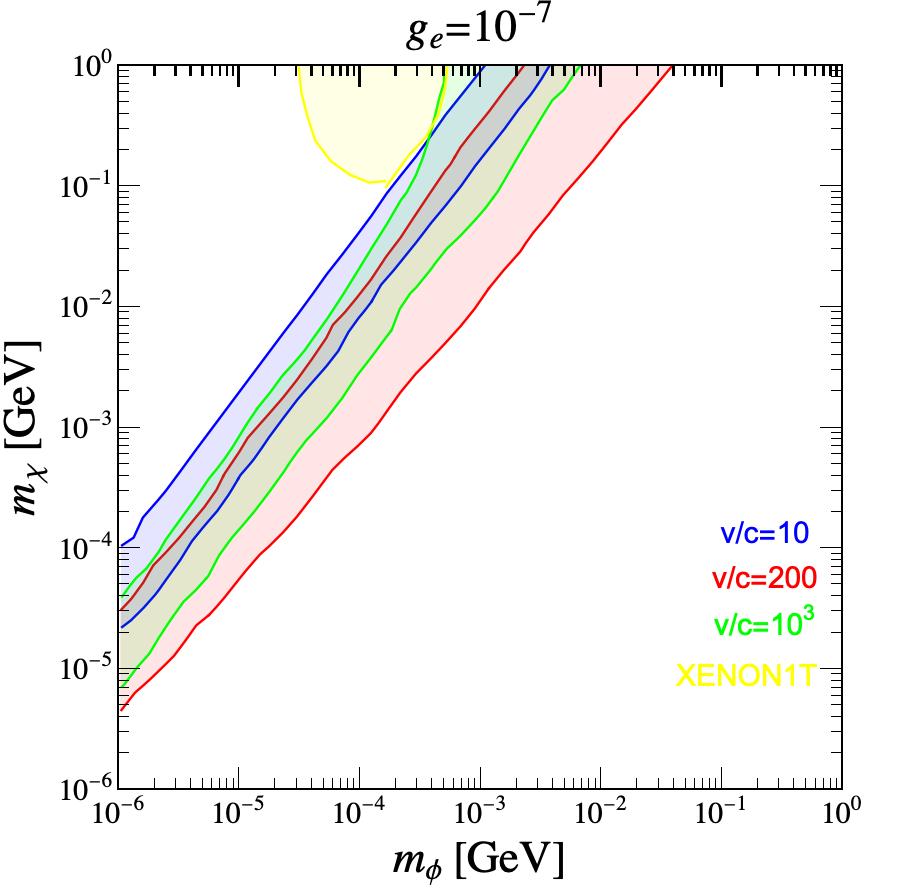}
	\includegraphics[width=6.3cm]{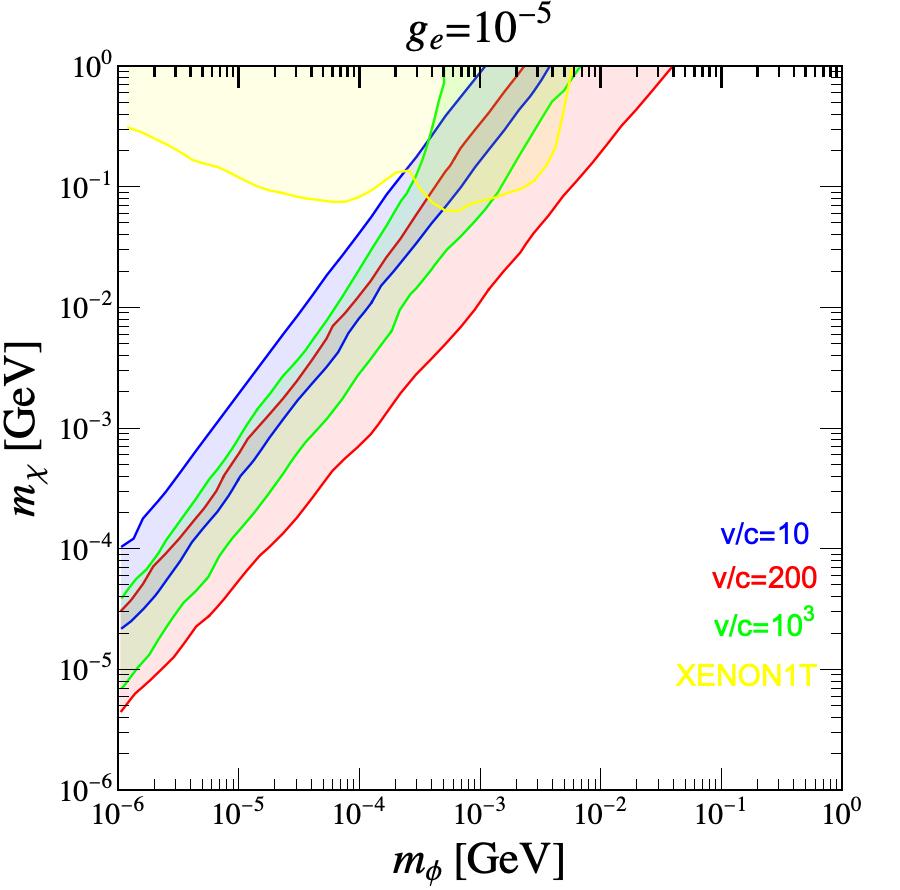}
   \vspace{-.5cm}
	\caption{Same as the right panle of Fig.\ref{fig2}, but for DM-electron scattering }
	\label{fig5}
\end{figure}

Following Refs.~\cite{Essig:2012yx,Essig:2017kqs,Wang:2023xgm} we can get the conversion from $E_{er}$ to electron yield $n_e$. Then we convert $n_e$ to the number of photoelectrons (PE). Finally, we scan over the masses of DM and the mediator. The constraints on the DM-electron scattering cross-section can be obtained by using the experimental data from XENON10 (15-kg-days) ~\cite{Essig:2012yx}, XENON100 (30-kg-years) ~\cite{XENON:2016jmt,XENON100:2011cza} and XENON1T (1.5 tones-years) ~\cite{XENON:2019gfn}. 

Fig. \ref{fig4} shows the XENON exclusion regions on the plane of DM-electron scattering cross-section versus the DM mass for the coupling $g_e=10^{-5}$ and $g_e=10^{-7}$ (note that the coupling $g_{\chi}$ is fixed by the DM relic density). 
Similar to the CRDM, in Fig. \ref{fig5} we show the regions allowed by solving the small-scale problems, compared with the XENON1T exclusion regions.
The difference between the left panel and right panel of Fig. \ref{fig4} comes from the mass of the mediator. In the left panel of Fig. \ref{fig4}, a smaller coupling $g_e$ requires a heavier mediator and the form factor is about 1.  The right panel of Fig. \ref{fig4} has a boarder exclusion area due to a lighter mediator required by a larger coupling $g_e$.  

\begin{figure}[ht]
	\centering
	\includegraphics[width=10cm]{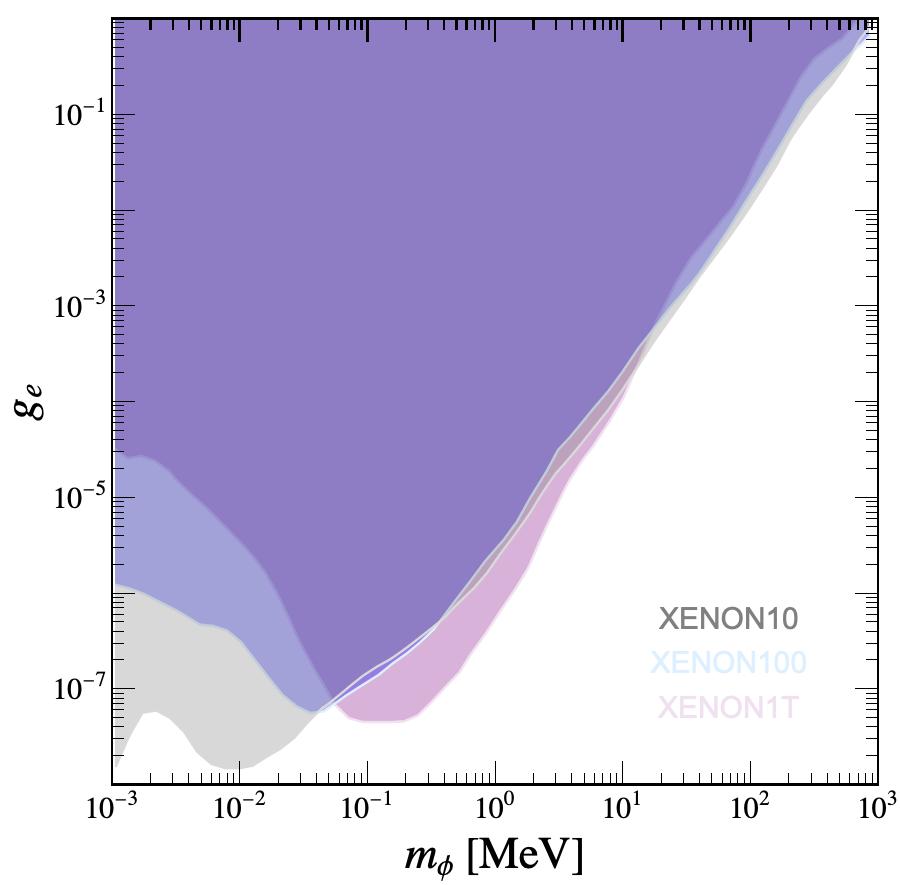}
  \vspace{-.5cm}
	\caption{The XENON exclusion regions on the plane of mediator-electron coupling versus the mediator mass.}
	\label{fig6}
\end{figure}

Finally we give the constraints on the mediator-electron coupling from Xenon1T, Xenon100 and Xenon10 data.
The results are shown in Fig. \ref{fig6}. We see that the coupling $g_e$ can be constrained down to $10^{-8}$ which corresponds to a mixing angle $\sin \theta \sim 10^{-3}$. Such a sizable mixing angle may also be constrained by the fifth force search, the beam dump experiment~\cite{Liu:2016qwd}, the stellar cooling from HB stars ~\cite{Hardy:2016kme} and the RG stars ~\cite{Hardy:2016kme}, or the SN1987A.  Anyway, even if other constraints are quite stringent, the direct detection can provide independent limits for such SIDM below GeV.  

\section{Conclusion}\label{sec5}
We studied the direct detection of the Higgs portal for a sub-GeV self-interacting dark matter with a scalar mediator,  for which the dark matter self-scattering cross-section is in the Born region.
Considering the constraints from CMB, BBN and the dark matter relic density, we considered the detection via two approaches: one is the cosmic-ray accelerated dark matter scattering off the nucleon, the other is the electron recoil caused by the halo dark matter. We presented direct detection limits for the parameter space of light dark matter and scalar mediator.  
We found that the detectability in either approach needs a sizable mediator-Higgs mixing angle ($\sin\theta$) which is larger than one for the CRDM approach and larger than $10^{-3}$ for the electron recoil approach. While the former case cannot be realized in the Higgs-portal light SIDM model with a scalar mediator, the latter case may also be constrained by some astrophysical observations or beam dump experiments. So we concluded that the direct detection experiments should explore such a sub-GeV self-interacting dark matter via CRDM or electron recoil (albeit rather challenging) to provide limits  independent of other limits from  astrophysical observations or beam dump experiments.

\section*{Acknowledgements}
This work was supported by the National
Natural Science Foundation of China (NNSFC) under grant Nos.  11821505, 12075300 and 12335005,
by the Research Fund for Outstanding Talents from Henan Normal University (5101029470335), by the Peng-Huan-Wu Theoretical Physics Innovation Center (12047503) funded by NNSFC. 


\bibliographystyle{apsrev}
\bibliography{note}

\end{document}